\documentclass[12pt,amsmath,amsfonts]{JHEP3}

\title{On Orbifold Compactification of ${\mathcal{N}}=2$ Supergravity in
Five Dimensions \\}
\author{
F. P. Zen$^a$, B. E. Gunara$^a$,  Arianto$^{a,b}$ and
H. Zainuddin$^c$\\

$^a$Theoretical Physics Laboratory, Department of Physics,\\
 Institute of Technology Bandung \\
Jl. Ganesha 10 Bandung 40132, Indonesia.\\

$^b$Department of Physics, Udayana University \\
Jl. Kampus Bukit Jimbaran Denpasar 80361, Indonesia. \\

$^c$Theoretical Studies Laboratory, ITMA, Universiti Putra Malaysia,\\
43400 UPM Serdang, Selangor, Malaysia. \\

E-mail: fpzen@fi.itb.ac.id, bobby@fi.itb.ac.id,
ariphys@student.fi.itb.ac.id, hisham@fsas.upm.edu.my.}

\preprint{}

\abstract{We study compactification of five dimensional ungauged
${\mathcal{N}}=2$ supergravity coupled to vector- and
hypermultiplets on orbifold $S^1/{\mathbb{Z}}_2$. In the model,
the vector multiplets scalar manifold is arbitrary while the
hypermultiplet scalars span a generalized self dual Einstein
manifold constructed by Calderbank and Pedersen. The bosonic and
the fermionic sectors of the low energy effective
${\mathcal{N}}=1$ supergravity in four dimensions are derived.}

\begin{document}

\section{Introduction}

Compactification of the five dimensional ${\mathcal{N}}=2$
supersymmetry on singular space $S^1/{\mathbb{Z}}_2$ has achieved
phenomenological interest since it provides ${\mathcal{N}}=1$
supersymmetry in four dimensions. Furthermore, four dimensional
${\mathcal{N}}=1$ vacua for the supersymmetric versions of the two
branes Randall-Sundrum scenario of the five dimensional
${\mathcal{N}}=2$ supergravity has been obtained in \cite{Bagg}.
In this scenario one places two 3-branes with opposite tension at
the orbifold $S^1/{\mathbb{Z}}_2$ which is the boundaries of
(4+1)-dimensional Anti de Sitter spacetime. The distance between
the branes is
set by the expectation value of a modulus field, called radion. Furthermore, starting from the simplest model, namely pure supergravity theory one can derive the effective ${\mathcal{N}}=1$ theory as it was shown in \cite{Bagg1}. \\
\indent A model consists of single hypermultiplet whose moduli
space of toric self-dual Einstein (TSDE) in five dimensional
${\mathcal{N}}=2$ supergravity has been studied to construct
domain wall solutions \cite{LC}. They investigated the associated
supersymmetric flows to prove the existence of domain walls which
admit Randall-Sundrum flows. However, in this model, the K\"{a}hler subspace of the TSDE is still unclear. \\
\indent The purpose of this paper is to obtain a four dimensional
${\mathcal{N}}=1$ theory via  $S^1/{\mathbb{Z}}_2$
compactification of the five dimensional ${\mathcal{N}}=2$
supergravity coupled to arbitrary vector multiplets and a
hypermultiplet which is chosen to be a generalized self dual
Einstein manifold admitting torus symmetry constructed by
Calderbank and Pedersen \cite{CP}.
Our aim is to find the K\"{a}hler subspace of toric self dual Einstein (TSDE) spaces. \\
\indent Our starting point is the five dimensional
${\mathcal{N}}=2$ supergravity coupled to arbitrary vector
multiplets  and a hypermultiplet. First, the theory can be
compactified down to four dimensions along an $S^1$ of radius $R$
parametrized by $x_5$, resulting in a nonchiral four dimensional
${\mathcal{N}}=2$ theory. However, we are interested in the chiral
four dimensional ${\mathcal{N}}=1$ theory. Second, to obtain a
chiral four dimensional ${\mathcal{N}}=1$ theory, we consider
compactification of the $x_5$ coordinate on the
$S^1/{\mathbb{Z}}_2$ orbifold. The ${\mathbb{Z}}_2$ action is as
usual $x_5 \rightarrow -x_5$ and two fixed points are at $x_5 =0$
and $x_5=\pi R$. We begin to mod $S^1$ by ${\mathbb{Z}}_2$. In
order the reduction to be consistent, we must first make a certain
parity assignment to the fields such that the Lagrangian is
invariant under $x_5 \rightarrow -x_5$. Then, when $S^1$ is modded
out by ${\mathbb{Z}}_2$, only the even parity fields survive on the two fixed points. \\
\indent The paper is organized as follows. Section 2 briefly
reviews the ungauged five dimensional ${\mathcal{N}}=2$
supergravity coupled to  vector- and hypermultiplets. We present
the action of the ungauged five dimensional ${\mathcal{N}}=2$
supergravity.  Section 3 presents a detailed derivation of the
compactification of five dimensional ${\mathcal{N}}=2$
supergravity coupled to vector multiplets and hypermultiplets.
First, we discuss some basic analysis of the $S^1/{\mathbb{Z}}_2$
orbifold compactification. The starting point is the bosonic
sector of the five dimensional ${\mathcal{N}}=2$ supergravity
theory. Second, after modding out ${\mathbb{Z}}_2$, the odd fields
are projected out and the surviving fields fit into multiplets of
the chiral four dimensional ${\mathcal{N}}=1$ supergravity. The
boundary action arising from dimensional reduction can be
constructed in a straightforward way and it is obtained by
truncating four dimensional ${\mathcal{N}}=1$ supergravity
according to the ${\mathbb{Z}}_2$ projection. We present then the
resulting of the compactification of the bosonic and the fermionic
sector. We conclude our results in section 4. Finally, in
Appendices A, B and C we summarize our notation, convention and
some of the detailed calculations.

\section{Ungauged five dimensional ${\mathcal{N}}=2$ supergravity}

This section describes supergravity theory in five dimensions with
${\mathcal{N}}=2$ supersymmetry in which a supergravity multiplet
is coupled to matter multiplets. The coupling to vector multiplets
was given in \cite{MGP1,MGP2} and the addition of tensor
multiplets was considered in \cite{GZ00}. Furthermore, the full
couplings of ${\mathcal{N}}=2$ supergravity theory in five
dimensions was constructed in \cite{CD00}.\footnote{See also
\cite{BCT} and references therein.}

\subsection{Pure gravitational multiplet}

Five dimensional gravitational multiplet consists of the metric
$\hat{g}_{\hat{\mu}\hat{\nu}}$, doublet symplectic Majorana
gravitinos $\psi^i_{\hat{\mu}}$ and a vector field
$\hat{A}_{\hat{\mu}}$ (graviphoton) \cite{EC}. The greek hatted
indices are five dimensional space-time indices and run over
values $0,\ldots,3,5$. The index $i$ of the gravitinos runs from 1
to 2.

The bosonic part of the action for the gravitational multiplet
takes the form:
\begin{equation}
S= -\frac{1}{2\kappa^2_5} \int d^5x \sqrt{-\hat{g}} \Bigg[
\hat{R}+ \hat{F}_{\hat{\mu}\hat{\nu}} \hat{F}^{\hat{\mu}\hat{\nu}}
+\frac{1}{6\sqrt{2}\sqrt{-\hat{g}}}\epsilon^{\hat{\mu}\hat{\nu}\hat{\rho}\hat{\sigma}\hat{\lambda}}
{\hat{F}}_{\hat{\mu}\hat{\nu}}
{\hat{F}}_{\hat{\rho}\hat{\sigma}}{\hat{A}}_{\hat{\lambda}}
\Bigg].
\end{equation}
Furthermore, one can also couple the pure ${\mathcal{N}}=2$
gravitational multiplet to arbitrary number of vector- and
hypermultiplets. This will be discussed in the next section.

\subsection{Couplings of vector- and hypermultiplets}

\subsubsection{The scalar manifold}
First, let us describe $n_V$ vector multiplets of
${\mathcal{N}}=2$ supergravity\footnote{We omit tensor multiplets
for simplicity. For ${\mathcal{N}}=2$ supergravity coupled to
tensor multiplets see \cite{GZ00}.}. We now have $n_V+1$ vector
fields $A_{\mu}^{I}$, $n_V$ symplectic pairs of gauginos
$\lambda^{ia}$, and $n_V$ real scalars $\phi^x$. It is convenient
to group vectors with the graviphoton so that the index
$I=0,1,\ldots,n_V$ and $a=1,\ldots,n_V$ are corresponding flat
space indices. The kinetic term of the scalars defines the sigma
model:
\begin{equation}
{\mathcal{L}}_{kin}=-\frac{1}{2}g_{xy}(\phi)\partial_{\hat{\mu}}\phi^x\partial^{\hat{\mu}}\phi^y.
\end{equation}
The vector multiplet scalars $\phi^x$, $x=1,\ldots,n_V$,
parametrize the target space $\mathcal{S}$ where $x$ represents
curved indices. The metric $g_{xy}$ can be interpreted as a metric
on a Riemannian manifold $\mathcal{S}$ called the very special
real geometry because it can be viewed as a hypersurface by an
$n_V$ polynomial of degree three
\begin{equation}
N(h)=C_{IJK}h^Ih^Jh^K =1 \;,
\end{equation}
where $h^I=h^I(\phi^x)$. Moreover, the gauge coupling of the theory can be expressed as
\begin{equation}
a_{IJ}(h)=-\frac{1}{3}\frac{\partial}{\partial h^I}\frac{\partial}{\partial h^J}lnN(h)|_{N=1}\;.
\end{equation}
Restricting to the submanifold  we can then write the  metric $g_{xy}(\phi)$ as:%
\begin{equation}
g_{xy}(\phi)=\frac{\partial h^I}{\partial \phi^x}\frac{\partial h^J}{\partial \phi^y}a_{IJ}(h).
\end{equation}
Secondly, we discuss $n_H$ hypermultiplets in which it contains $4n_H$ real scalars $q^X$ and $2n_H$ symplectic Majorana fermions (hyperinos) $\zeta^A$ where $X=1,\ldots,4n_H$ and $A=1,\ldots,2n_H$. As in the previous case, the central object is the metric $g_{XY}$ of the sigma model:
\begin{equation}
{\mathcal{L}}_{kin}=-\frac{1}{2}g_{XY}(q)\partial_{\hat{\mu}}q^X\partial^{\hat{\mu}}q^Y.
\end{equation}
Again, $g_{XY}$ is the metric on a Riemannian manifold
${\mathcal{Q}}$ on which the scalars $q^X$ are the coordinates and
thus $X=1,\ldots, 4n_H$ are the curved indices labelling the
coordinates. Local supersymmetry further implies that
$\mathcal{Q}$
has to be a quaternionic K\"{a}hler manifold \cite{BW}.\\
\indent Thus from the above discussion it shows that the scalar
manifold $\mathcal{M}$ is a direct product of a very special
manifold ${\mathcal{S}}$ and a quaternionic manifold
$\mathcal{Q}$:
\begin{equation}
{\mathcal{M}}={\mathcal{S}}\otimes {\mathcal{Q}},
\end{equation}
with $\phi^{x}\in{\mathcal{S}}$, $q^X\in{\mathcal{Q}}$. \\
\indent We now write the action of ${\mathcal{N}}=2$ supergravity
which is needed for our analysis\footnote{For complete action see
\cite{BCT}.}
\begin{eqnarray}
\label{action} S&=& \int d^5x \sqrt{-\hat{g}}
\Bigg[\frac{1}{2\kappa^2_5} \hat{R} - \frac{1}{4} a_{IJ}
\hat{F}^I_{\hat{\mu}\hat{\nu}} \hat{F}^{J \hat{\mu}\hat{\nu}} -
\frac{1}{2} g_{xy}\partial_{\hat{\mu}}\phi^x
\partial^{\hat{\mu}}\phi^y
- \frac{1}{2} g_{XY}\partial_{\hat{\mu}}q^X \partial^{\hat{\mu}}q^Y \nonumber\\
& &-\frac{1}{2\kappa^2_5} \bar{\psi}_{\hat{\rho}} \gamma^{\hat{\rho}\hat{\mu}\hat{\nu}} D_{\hat{\mu}} \psi_{\hat{\nu}}
- \frac{1}{2} \bar{\lambda}_x \gamma^{\hat{\mu}} D_{\hat{\mu}} \lambda^x
- \bar{\zeta}^A \gamma^{\hat{\mu}} D_{\hat{\mu}} \zeta_A \nonumber\\
& & +\frac{1}{6\sqrt{6}\sqrt{-\hat{g}}}\epsilon^{\hat{\mu}\hat{\nu}\hat{\rho}\hat{\sigma}\hat{\lambda}}
C_{IJK} \hat{F}^I_{\hat{\mu}\hat{\nu}} \hat{F}^J_{\hat{\rho}\hat{\sigma}} \hat{A}^K_{\hat{\lambda}}
 - \frac{i\sqrt{6}}{16\kappa_5} h_{I} \hat{F}^{\hat{c}\hat{d}I} \bar{\psi}^{\hat{a}} \gamma_{\hat{a}\hat{b}\hat{c}\hat{d}} \psi^{\hat{b}} \nonumber\\
& &+ \frac{i\kappa_5}{4} \sqrt{\frac{2}{3}} \Bigg(\frac{1}{4}
g_{xy} h_I + T_{xyz} h^z_I \Bigg) \lambda^x
\gamma^{\hat{a}\hat{b}} \hat{F}^{I}_{\hat{a}\hat{b}} \lambda^y +
\frac{i\kappa_5}{8}\sqrt{6}h_I\bar{\zeta}_A\gamma^{\hat{a}\hat{b}}
\hat{F}^I_{\hat{a}\hat{b}}\zeta^A \Bigg],
\end{eqnarray}
where the covariant derivatives are given by
\begin{eqnarray}
D_{\hat{\mu}} \lambda^{xi} &=& \partial_{\hat{\mu}}
\lambda^{xi}+\partial_{\hat{\mu}} \phi^y \Gamma_{yz}^x
\lambda^{zi} +\frac{1}{4} \omega_{\hat{\mu}}{}^{\hat{a}\hat{b}}
\gamma_{\hat{a}\hat{b}}\lambda^{xi} +\partial_{\hat{\mu}} q^X
\omega_{Xj}{}^i\lambda^{xj},\\  \label{cov1} D_{\hat{\mu}}
\zeta^{A} &=& \partial_{\hat{\mu}} \zeta^{A} +
\partial_{\hat{\mu}} q^{X} \omega_{{X}{B}}{}^{A} \zeta^{B}
+\frac{1}{4} \omega_{\hat{\mu}}{}^{\hat{b}\hat{c}}
\gamma_{\hat{b}\hat{c}} \zeta^{ A}, \\ \label{cov2} D_{\hat{\mu}}
\psi^i_{\hat{\nu}} &=& \bigg(\partial_{\hat{\mu}} + \frac{1}{4}
\omega_{\hat{\mu}}{}^{\hat{b}\hat{c}}
\gamma_{\hat{b}\hat{c}}\bigg)\psi_{\hat{\nu}}^i-\partial_{\hat{\mu}}
q^X \omega_{X}{}^{ij}\psi_{\hat{\nu}j}. \label{cov3}
\end{eqnarray}
\subsubsection{Toric Self Dual Einstein Spaces}
Let us now consider the hypermultiplet sector that is four
dimensional $(n_H=1)$ which has self dual property beside Einstein
spaces. Our choice is the most general space admitting $T^2$
isometry which has been shown in \cite{CP}. The metric has the
form
\begin{eqnarray}
ds^2 &=& -\Bigg[ \frac{1}{4{\rho}^2} -
\frac{(f^2_{\rho}+f^2_{\eta})}{f^2}\Bigg] \bigg( d{\rho}^2 +
d{\eta}^2 \bigg)
\nonumber\\
&& -\frac{\bigg( (f - 2\rho f_{\rho})\alpha - 2\rho f_{\eta}\beta
\bigg)^2 +\bigg( -2\rho f_{\eta}\alpha+(f+2\rho f_{\rho})\beta
\bigg)^2} {f^2 \bigg( f^2 - 4{\rho}^2(f^2_{\rho} + f^2_{\eta})
\bigg)}, \label{TSD}
\end{eqnarray}
where $\alpha=\sqrt{\rho}d\phi$ and $\beta=(d\psi + \eta d\phi)$.
The function $f(\rho,\eta)$ satisfies the Laplace equation in two dimensional hyperbolic space spanned by $(\rho,\eta)$
\begin{equation}\label{diff:eps}
\rho^2 (f_{\rho\rho}+f_{\eta\eta})=\frac{3}{4}f,
\end{equation}
with $f_{\rho\rho}=\frac{{\partial}^2{f}}{\partial{\rho}^2}$ and $f_{\eta\eta}=\frac{{\partial}^2{f}}{\partial{\eta}^2}$. One takes $\rho>0$ and $\eta \in {\mathbb{R}}$ while $(\phi,\psi)$ are periodic coordinates. Furthermore, it has positive scalar curvature if $f$ satisfies $f^2 > 4{\rho}^2(f^2_{\rho}+f^2_{\eta})$ and negative if $f^2 < 4{\rho}^2(f^2_{\rho}+f^2_{\eta})$.

\section{Compactification on the orbifold $S^1/{\mathbb{Z}}_2$}

\subsection{Analysis of the orbifold transformation}
Now we turn our attention to  consider the five dimensional supergravity on the orbifold $S^1/{\mathbb{Z}}_2$ \footnote{We discuss $S^1$ compactification in Appendix B.}.  As we shall see that the five dimensional ${\mathcal{N}}=2$ supergravity is reduced to four dimensional ${\mathcal{N}}=1$ supergravity.\\
\indent There are two ways to employ the  compactification on
orbifold. In the so-called downstair approach, we consider the
five dimensional supergravity with $x^5\in
S^1/{\mathbb{Z}}_2=[0,\pi R]$, so $x^5\sim x^5+2\pi R$ and
$x^5\sim -x^5$. Then, the five dimensional space-time is
$M^5_{down}=M^4\times (S^1/{\mathbb{Z}}_2)=M^4\times[0,\pi R]$.
Locally, we still have ordinary the five dimensional supergravity,
but on boundary special thing can occur. The boundary consist of
two four dimensional space-time, one at $x^5=0$ called $M^4$ and
one at
$x^5=\pi R$ called ${M'}^4$.\\
\indent In the so called upstair approach, we consider the five
dimensional supergravity with $x^5$ on $S^1$ so that $x^5\sim
x^5+2\pi R$ and the five dimensional space-time is $M^5_{up} = M^4
\times S^1$. Then, we define the orbifold transformation
${\mathcal{O}}$ by
\begin{equation}
{\mathcal{O}}: {x^5}{\rightarrow} {-x^5}, \quad {x^{\mu}} {\rightarrow} {x^{\mu}}.
\end{equation}
The fixed points of ${\mathcal{O}}$ are the points with $x^5=0$ or
$x^5=\pi R$, so the fixed points consist of two four dimensional
space-time $M^4$ and ${M'}^4$ which are the boundary of
$M^5_{up}$. To get the same picture as in the downstair approach,
we demand that the physics should be invariant under the orbifold
transformation $\mathcal{O}$. We must in particular have the
distance $d\hat{s}^2$ invariant,
\begin{eqnarray}
d\hat{s}^2&=&\hat{g}_{\hat{\mu}\hat{\nu}}dx^{\hat{\mu}}dx^{\hat{\nu}} \nonumber\\
&=&\hat{g}_{\mu\nu}dx^\mu dx^\nu +2\hat{g}_{\mu 5}dx^\mu dx^5+ \hat{g}_{55} dx^5 dx^5 \nonumber\\
&\rightarrow& \hat{g}_{\mu\nu}dx^\mu dx^\nu +2\hat{g}_{\mu
5}dx^\mu (-dx^5)+\hat{g}_{55}(-dx^5)(-dx^5).
\end{eqnarray}
The invariant properties of the distance require that
\begin{equation}
\hat{g}_{\mu\nu}\rightarrow \hat{g}_{\mu\nu}, \quad \hat{g}_{\mu 5}\rightarrow -\hat{g}_{\mu 5}, \quad
\hat{g}_{55}\rightarrow \hat{g}_{55},
\end{equation}
such that on $M^4$ and ${M'}^4$, we have that $\hat{g}_{\mu 5}=0$.\\
\indent Next, we analyze the gravitational part of the five
dimensional supergravity action. Under parity transformation, the
action become
\begin{equation}
\int d^5x\sqrt{-\hat{g}}\hat{R}\rightarrow \int dx^4(-dx^5)\sqrt{-\hat{g}}\hat{R}=-\int d^5x\sqrt{-\hat{g}}\hat{R},
\end{equation}
where $R$ and $\sqrt{-g}$ are invariant under $x^5 \rightarrow -x^5$. Therefore, we need that the other terms change sign as well.\\
\indent The transformation of $\hat{F}_{\mu\nu}^I$ can be seen
directly from the transformation of $\hat{g}^{\mu 5}$
\begin{eqnarray}
\hat{F}_{\mu\nu}^I&=&\partial_\mu \hat{A}^I_\nu-\partial_\nu \hat{A}^I_\mu \rightarrow -\hat{F}^I_{\mu\nu}, \nonumber\\
\hat{F}_{\mu 5}^I&=&\partial_\mu \hat{A}^I_5-\partial_5 \hat{A}^I_\mu \rightarrow \hat{F}^I_{\mu 5},
\end{eqnarray}
from which it can be checked that FF-term changes sign:
\begin{eqnarray}
S_{FF}&=&\int d^5x\sqrt{-g}a_{IJ}\hat{F}^I_{\hat{\mu}\hat{\nu}}\hat{F}^{\hat{\mu}\hat{\nu} J} \nonumber\\
&=&\int d^5x\sqrt{-\hat{g}}a_{IJ}\hat{F}^I_{\hat{\mu}\hat{\nu}}\hat{F}^J_{\hat{\rho}\hat{\sigma}}
\hat{g}^{\hat{\mu}\hat{\rho}}\hat{g}^{\hat{\nu}\hat{\sigma}} \nonumber\\
&\rightarrow&\int(-d^5x)\sqrt{-\hat{g}}a_{IJ}(-\hat{F}^I_{\mu\nu})(-\hat{F}^J_{\rho\sigma})
\hat{g}^{\mu\rho}\hat{g}^{\nu\sigma} \nonumber\\
& &+2\int(-d^5x)\sqrt{-\hat{g}}a_{IJ}(\hat{F}^I_{\mu 5})(\hat{F}^J_{\nu 5})
(-\hat{g}^{\mu 5})(-\hat{g}^{\nu 5})\nonumber\\
&=&-\int d^5x\sqrt{-\hat{g}}a_{IJ}\hat{F}^I_{\hat{\mu}\hat{\nu}}\hat{F}^{\hat{\mu}\hat{\nu} J}.
\end{eqnarray}
Finally, we have $\hat{A}^I_\mu \rightarrow -\hat{A}^I_\mu$ and
$\hat{A}^I_5 \rightarrow \hat{A}^I_5$.\\
\indent Since the orbifold $S^1/{\mathbb{Z}}_2$ has boundaries at
the two fixed points, we have to add the extra terms to the action
$(\ref{action})$ and then get the modified action. First of all,
we derive the equation of motion including variation of Ricci
tensor. In other words, the derivative of the variation of the
metric is not zero on the boundary. In the following we only
consider variation of Ricci scalar which give the modified action
of the five dimensional supergravity theory on orbifold, and then
obtain the modified action of the five dimensional supergravity
which can be written as
\begin{equation}\label{eqmod:eps}
S_{5d}^{mod}=S+2\int_{\partial\Sigma}\hat{E},
\end{equation}
where $\hat{E}=\hat{h}^{\hat{\mu}}_{\phantom{\hat{\mu}}\hat{\nu}}\nabla_{\hat{\mu}}\hat{N}^{\hat{\nu}}$ is the
trace of the extrinsic curvature. In next section we discuss the second term
of the above equation.\\
\indent The variation of the five dimensional Ricci scalar is given by
\begin{eqnarray}\label{eqvar:eps}
\delta \hat{S}&=&\int d^5x \Bigg[\delta\sqrt{-\hat{g}}\hat{R}
+\sqrt{-\hat{g}}\delta
\hat{g}^{\hat{\mu}\hat{\nu}}\hat{R}_{\hat{\mu}\hat{\nu}}
+\sqrt{-\hat{g}}\hat{g}^{\hat{\mu}\hat{\nu}}\delta \hat{R}_{\hat{\mu}\hat{\nu}}\Bigg] \nonumber\\
&=&\int d^5x
\sqrt{-\hat{g}}\Bigg[g\frac{1}{2}\hat{g}^{\hat{\rho}\hat{\sigma}}\hat{R}
-\hat{g}^{\hat{\mu}\hat{\rho}}\hat{g}^{\hat{\sigma}\hat{\nu}}\hat{R}_{\hat{\mu}\hat{\nu}}\Bigg]\delta \hat{g}_{\hat{\rho}\hat{\sigma}} \nonumber\\
& &+\int d^5x \sqrt{-\hat{g}}\nabla^{\hat{\rho}}
\Bigg[-\nabla_{\hat{\rho}}(\hat{g}^{\hat{\mu}\hat{\nu}}\delta\hat{g}_{\hat{\mu}\hat{\nu}})
+\nabla^{\hat{\mu}}\delta \hat{g}_{\hat{\rho}\hat{\mu}}\Bigg],
\end{eqnarray}
where we have used
\begin{equation}\label{pal}
\delta
\hat{R}_{\hat{\mu}\hat{\nu}}=\frac{1}{2}\hat{g}^{\hat{\rho}\hat{\sigma}}
\Big(-\nabla_{\hat{\mu}}\nabla_{\hat{\nu}} (\delta
\hat{g}_{\hat{\rho}\hat{\sigma}})
+\nabla_{\hat{\rho}}\nabla_{\hat{\nu}}(\delta
\hat{g}_{\hat{\sigma}\hat{\mu}})
+\nabla_{\hat{\rho}}\nabla_{\hat{\mu}}(\delta
\hat{g}_{\hat{\sigma}\hat{\nu}})
-\nabla_{\hat{\rho}}\nabla_{\hat{\sigma}}(\delta
\hat{g}_{\hat{\mu}\hat{\nu}})\Big),
\end{equation}
and the metric postulate $\nabla_{\hat{\rho}}g_{\hat{\mu}\hat{\nu}}=0$.
After plugging $(\ref{pal})$ into $(\ref{eqvar:eps})$, finally we get
\begin{eqnarray} \label{eqvar0:eps}
\delta \hat{S}&=&\int d^5x \sqrt{-\hat{g}}
\Bigg[\frac{1}{2}\hat{g}^{\hat{\rho}\hat{\sigma}}\hat{R}
-\hat{R}^{\hat{\rho}\hat{\sigma}}\Bigg]\delta
\hat{g}_{\hat{\rho}\hat{\sigma}} +\int d^5x
\sqrt{-\hat{g}}\nabla^{\hat{\mu}}\hat{T}_{\hat{\mu}},
\end{eqnarray}
where
\begin{equation}
\hat{T}_{\hat{\mu}}=\hat{g}^{\hat{\nu}\hat{\rho}}(\nabla_{\hat{\rho}}\delta \hat{g}_{\hat{\mu}\hat{\nu}}
-\nabla_{\hat{\mu}}\delta \hat{g}_{\hat{\nu}\hat{\rho}}).
\end{equation}

Using Gauss theorem, we can rewrite the last term of
$(\ref{eqvar0:eps})$:
\begin{equation}\label{eqbo:eps}
\int_{\Sigma} d^5x \partial^{\hat{\mu}}(\sqrt{-\hat{g}}\hat{T}_{\hat{\mu}})=
\int_{\partial\Sigma} d^4x\sqrt{-\hat{g}}\hat{N}^{\hat{\mu}}\hat{T}_{\hat{\mu}},
\end{equation}
where $\hat{N}^{\hat{\mu}}$ is the normal to the surface. This
term is zero because the derivative of the variation of the metric
is zero on the boundary.
This term modifies the action of the five dimensional supergravity theory on orbifold . \\
\indent Next, after assuming $\delta \hat{g}_{\hat{\mu}\hat{\nu}}$
is a constant on the surface $\partial{\Sigma}$, then the equation
$(\ref{eqbo:eps})$ can be written down as:
\begin{equation}\label{eqbo0:eps}
\int_{\Sigma} d^5x \partial^{\hat{\mu}}(\sqrt{-\hat{g}}\hat{T}_{\hat{\mu}})=
-\int_{\partial\Sigma} d^4x\sqrt{-\hat{g}} \hat{N}^{\hat{\mu}}\hat{h}^{\hat{\nu}\hat{\rho}}\nabla_{\hat{\mu}}(\delta\hat{g}_{\hat{\nu}\hat{\rho}}),
\end{equation}
where we have used
$\hat{h}^{\hat{\nu}\hat{\rho}}\nabla_{\hat{\rho}}\delta
\hat{g}_{\hat{\mu}\hat{\nu}}=0$. The boundaries can be considered
as a four dimensional surface embedded in the five dimensional
space-time.

We now define a quantity $\hat{E}$, whose variation equals to the
equation $(\ref{eqbo0:eps})$. We have
\begin{equation}
\hat{E}_{\hat{\mu}\hat{\nu}}=\hat{h}_{\hat{\mu}}^{\phantom{\hat{\mu}}\hat{\rho}}\nabla_{\hat{\rho}}\hat{N}_{\hat{\nu}}, \quad
\hat{h}_{\hat{\mu}\hat{\nu}}=\hat{g}_{\hat{\mu}\hat{\nu}}-\hat{N}_{\hat{\mu}}\hat{N}_{\hat{\nu}},
\end{equation}
where
\begin{equation}
\hat{N}_{\hat{\mu}}=\pm \delta^5_{\hat{\mu}}\sqrt{\hat{g}_{55}}.
\end{equation}
From the above equation we see that induced metric on the surface
$\partial{\Sigma}$ has
\begin{equation}
\hat{h}_{55}=0.
\end{equation}
The trace of extrinsic curvature $\hat{E}$ is calculated according
to
\begin{eqnarray}\label{eqtr:eps}
\hat{E}&=&\hat{g}^{\hat{\mu}\hat{\nu}}\hat{E}_{\hat{\mu}\hat{\nu}}=
\hat{g}^{\hat{\mu}\hat{\nu}}\hat{h}_{\hat{\mu}}^{\phantom{\hat{\mu}}\hat{\rho}}\nabla_{\hat{\rho}}\hat{N}_{\hat{\nu}} \nonumber\\
&=&\hat{h}^{\hat{\nu}\hat{\rho}}(\partial_{\hat{\rho}}\hat{N}_{\hat{\nu}}-\hat{\Gamma}^{\hat{\sigma}}_{\hat{\rho}\hat{\nu}}\hat{N}_{\hat{\sigma}})
=\hat{h}^{5\hat{\rho}}(\partial_{\hat{\rho}}\hat{N}_5-\hat{\Gamma}^{5}_{\hat{\rho}\hat{\nu}}\hat{N}_5) \nonumber\\
&=&\hat{h}^{5\hat{\rho}}\partial_{\hat{\rho}}\hat{N}_5
-\frac{1}{2}\hat{g}^{5\hat{\lambda}}\hat{h}^{\hat{\nu}\hat{\rho}}\big[\partial_{\hat{\rho}}\hat{g}_{\hat{\lambda}\hat{\nu}}
-\partial_{\hat{\nu}}\hat{g}_{\hat{\lambda}\hat{\rho}}
+\partial_{\hat{\lambda}}\hat{g}_{\hat{\nu}\hat{\rho}}\big]\hat{N}_5,
\end{eqnarray}
where we have used definition of the Christoffel symbol.\\
\indent Now we use the fact that both $\hat{g}_{\hat{\mu}\hat{\nu}}$ and
$\hat{h}_{\hat{\mu}\hat{\nu}}$ are block diagonal
\begin{eqnarray}
\hat{E}&=&\hat{h}^{55}\partial_{5}\hat{N}_5
-\frac{1}{2}\hat{g}^{55}\hat{h}^{5\hat{\rho}}\partial_{\hat{\rho}}\hat{g}_{55}\nonumber\\
& &-\frac{1}{2}\hat{g}^{55}\hat{h}^{\hat{\nu}5}\partial_{\hat{\nu}}\hat{g}_{55}
+\frac{1}{2}\hat{g}^{55}\hat{h}^{\hat{\nu}\hat{rho}}\partial_{5}\hat{g}_{\hat{\nu}\hat{\rho}}\hat{N}_5 \nonumber\\
&=&\frac{1}{2}\hat{g}^{55}\hat{h}^{\hat{\nu}\hat{\rho}}\partial_{5}\hat{g}_{\hat{\nu}\hat{\rho}}\hat{N}_5,
\end{eqnarray}
where we have used the fact that
$\hat{h}^{\hat{\nu}5}\partial_{\hat{\nu}}\hat{g}_{55}=0$ since
this is a derivative along the surface and that the variation is
zero on the surface. The second term of $(\ref{eqmod:eps})$ comes
from variation of the extrinsic curvature $(\ref{eqtr:eps})$, we
get
\begin{equation}
\delta\hat{E}=\frac{1}{2}\hat{N}^{\hat{\mu}}\hat{h}^{\hat{\nu}\hat{\rho}}(\nabla_{\hat{\mu}}\delta\hat{g}_{\hat{\nu}\hat{\rho}}),
\end{equation}
which is identical to $(\ref{eqbo0:eps})$ apart from the factor
two. Here we are working in the massless sector of the theory,
where $\partial_5\hat{g}_{\hat{\mu}\hat{\nu}}=0$ so the second
term of $(\ref{eqmod:eps})$ does not contribute to our action.
However if we are working in the massive sector, this term can not
be ignored. In the following section we only consider the massless
sector to
obtain the low-energy  effective action of the four dimensional ${\mathcal{N}}=1$ supersymmetry theory. \\
\indent From the bosonic sector analysis, the five dimensional
fields can be even ($\Phi(x^5) = \Phi(-x^5)$) or odd ($\Phi(x^5) =
- \Phi(-x^5)$) under orbifold transformation. Note that the odd
fields must either vanish or be discontinuous at the fixed points,
hence they are not dynamical fields on the submanifolds.
Identifying periodicity of the scalar fields in $(\ref{TSD})$,
$\Phi \rightarrow \Phi + \Phi_0$ with $\Phi \equiv (\rho, \eta,
\psi, \phi)$
and under orbifold transformation $\Phi \rightarrow -\Phi$, we define the fields $(\rho, \eta)$ are even and $(\psi,\phi)$ are odd.\footnote{The odd parity of $(\rho,\eta)$ is not satisfied in the function $F(\rho, \eta)$, for example, for $F(\rho, \eta)= \frac{\sqrt{\rho^2 + (\eta - y)^2}}{\sqrt{\rho}}$. See $\cite{CP}$.}\\
\indent The action of $Z_2$ on the fermion fields and on the
spinor parameter $\epsilon$ of the supersymmetry transformations
is defined as \cite{ERA, YIN}:
\begin{eqnarray}
\psi^i_{\mu}(x^5) &=& \mathbf{P}(\psi )\gamma _5 M(q)^i{}_j \psi_{\mu}^j(-x^5)\,, \\
\lambda ^i(x^5)&=& \mathbf{P}(\lambda )\gamma _5 M(q)^i{}_j \lambda ^j(-x^5)\,,\\
\epsilon^i(x^5)&=&  \mathbf{P}(\epsilon)\gamma _5 M(q)^i{}_j \varepsilon ^j(-x^5),
\end{eqnarray}
where
\begin{equation}
M^i{}_j=m_1(q) (\sigma _1)^i{}_j + m_2(q) (\sigma _2)^i{}_j + m_3(q) (\sigma_3)^i{}_j\,,
\end{equation}
with $m_1, m_2, m_3\in$ real functions of $q$ and
$\mathbf{P}(\Psi)=\pm$, $\Psi\equiv(\lambda,\psi,\varepsilon)$.
Decomposing the five dimensional spinor $\Psi$, and its conjugate
$\bar{\Psi}$, into four-dimensional spinor, and following the
convention in \cite{Bagg1} it is given by
\begin{equation}
\Psi_{\hat{\mu}} =
\left(
\begin{array}{c}
\psi^1_{\hat{\mu} L}
\\
\phantom{bla}
\\
\overline{\psi}_{2\hat{\mu} R}
\end{array}
\right),
\end{equation}
and
\begin{equation}
\gamma^5   =
\left(
\begin{array}{ccc}
-i  & {} &  0
\\
\phantom{bla}
\\
0  & {} & i
\end{array}
\right), \qquad
\gamma^{\hat{a}} =
\left(
\begin{array}{cc}
0 & \sigma^{\hat{a}}
\\
\phantom{bla}
\\
\sigma^{\hat{a}} & 0
\end{array}
\right).
\end{equation}
\indent To complete the results  of this section, we must first
take certain parity assignments to the fields such that the action
stays invariant under $x^5 \rightarrow -x^5$. Then, when we mod
out ${\mathbb{Z}}_2$, only fields of even parity survive on the
two orbifold fixed points. The even parity fields are given by
\begin{eqnarray}
\hat{g}_{\mu\nu}, \quad \hat{g}_{55}, \hat{A}^I_5, \quad \rho, \quad \eta,\quad \psi^{+}_{\mu}, \quad
\psi^{-}_{5},\quad \zeta^1, \quad \lambda^1_x, \quad \epsilon^{+},
\end{eqnarray}
and those of odd parity are given by
\begin{eqnarray}
\hat{g}_{\mu 5}, \quad \hat{g}_{5 \mu}, \hat{A}^I_{\mu}, \quad \phi, \quad \psi, \quad \psi^{-}_{\mu}, \quad
\psi^{+}_{5},\quad \zeta^2, \quad \lambda^2_x, \quad \epsilon^{-},
\end{eqnarray}
where we define $\psi^{\pm}_{\hat{\mu}}=\frac{1}{\sqrt{2}} ( \psi^{1}_{\hat{\mu}} \pm \psi^{2}_{\hat{\mu}} )$.
\subsection{The bosonic sector}
In this subsection we derive the low energy effective ${\mathcal{N}}=1$ action via compactification of the bosonic part of the action of the ungauged ${\mathcal{N}}=2$ supergravity in five dimensions ($\ref{action}$) on the orbifold $S^1/{\mathbb{Z}}_2$ using the analysis above. \\
\indent Under the ${\mathbb{Z}}_2$ symmetry, the bosonic fields
($\hat{g}_{\mu\nu}, \hat{g}_{55}, \hat{A}^I_5, \rho, \eta$) have
to be even, while ($\hat{g}_{\mu 5}, \hat{A}^I_\mu, \phi, \psi$)
are odd with respect to the orbifold transformation. The analysis
of the orbifold transformation above are similar to $S^1$
compactification,\footnote{See Appendix B.} however all odd fields
are ruled out. Using the  ansatz,
\begin{equation}
d\hat{s}^2 = A(x)g_{\mu\nu}dx^{\mu}dx^{\nu}+B(x)dz^2,
\end{equation}
we find
\begin{eqnarray}\label{orbifold1:eps}
S_{S^1/{\mathbb{Z}}_2} &=& \int d^4 \sqrt{-g} \Bigg[
\frac{1}{2\kappa^2_4} \bigg( AB^{1/2}R + \frac{3}{2} A^{-1}B^{1/2}
\partial_{\mu} A \partial^{\mu} A
+ \frac{3}{2} B^{1/2} \partial_{\mu} A \partial^{\mu} B \bigg) \nonumber\\
& & - \frac{1}{2} \Bigg(\frac{\kappa^2_5}{\kappa^2_4}\Bigg)
AB^{-1/2} a_{IJ} \partial_{\mu} A^I_5 \partial^{\mu} A^I_5
- \frac{1}{2} \Bigg(\frac{\kappa^2_5}{\kappa^2_4}\Bigg) AB^{1/2} g_{xy} \partial_{\mu} \phi^x \partial^{\mu} \phi^y \nonumber\\
& & + \frac{1}{2} \Bigg(\frac{\kappa^2_5}{\kappa^2_4}\Bigg)
AB^{1/2} \Bigg( \frac{1}{4\rho^2}- \frac{f^2_\rho + f^2_\eta}{f^2}
\Bigg) \bigg( \partial_{\mu} \rho \partial^{\mu} \rho +
\partial_{\mu} \eta \partial^{\mu} \eta \bigg) \Bigg].
\end{eqnarray}
To put the equation $(\ref{orbifold1:eps})$ back into the
canonical form, we perform a Weyl rescaling of the metric which is
given by:
\begin{equation}
g_{\mu\nu} \rightarrow \tilde{g}_{\mu\nu}=A^{-1}B^{-1/2}g_{\mu\nu},
\quad g^{\mu\nu} \rightarrow \tilde{g}^{\mu\nu}=AB^{1/2}g^{\mu\nu}.
\end{equation}
So under this rescaling the equation $(\ref{orbifold1:eps})$ becomes:
\begin{eqnarray}\label{orbifold2:eps}
S_{S^1/{\mathbb{Z}}_2} &=& \int d^4 \sqrt{-g} \Bigg[
\frac{1}{2\kappa^2_4} \bigg(R - \frac{3}{8}  \partial_{\mu} ln B
\partial^{\mu} ln B \bigg) \nonumber\\
&& - \frac{1}{2} \Bigg(\frac{\kappa^2_5}{\kappa^2_4}\Bigg) B^{-1}
a_{IJ} \partial_{\mu} A^I_5 \partial^{\mu} A^I_5  - \frac{1}{2}
\Bigg(\frac{\kappa^2_5}{\kappa^2_4}\Bigg) g_{xy}
\partial_{\mu} \phi^x \partial^{\mu} \phi^y \nonumber\\
&& + \frac{1}{2} \Bigg(\frac{\kappa^2_5}{\kappa^2_4}\Bigg) \Bigg(
\frac{1}{4\rho^2}- \frac{f^2_\rho + f^2_\eta}{f^2} \Bigg) \bigg(
\partial_{\mu} \rho \partial^{\mu} \rho + \partial_{\mu} \eta
\partial^{\mu} \eta \bigg) \Bigg].
\end{eqnarray}
The four dimensional gravitational constant ($1/{\kappa^2_4}$)  can be expressed in terms of its
five dimensional counterpart as:
\begin{equation}
\kappa^2_4=\frac{\kappa^2_5}{2\pi R}.
\end{equation}
The equation (\ref{orbifold2:eps}) can be rewritten in the form:
\begin{eqnarray}\label{orbifold3:eps}
S_{S^1/{\mathbb{Z}}_2} &=&  \frac{1}{2\kappa^2_4} \int d^4
\sqrt{-g} \Bigg[ R - \frac{1}{\kappa^2_5} \hat{a}_{IJ}
\partial_{\mu} \hat{h}^I \partial^{\mu} \hat{h}^J
- \frac{2}{3} \hat{a}_{IJ} \partial_{\mu} A^I_5 \partial^{\mu} A^I_5 \nonumber\\
& &  + \kappa^2_5 \Bigg( \frac{1}{4\rho^2}- \frac{f^2_\rho +
f^2_\eta}{f^2} \Bigg) \bigg( \partial_{\mu} \rho \partial^{\mu}
\rho + \partial_{\mu} \eta \partial^{\mu} \eta \bigg) \Bigg],
\end{eqnarray}
where $\hat{a}_{IJ}=\frac{3\kappa^2_5}{2}B^{-1}a_{IJ}$, $\hat{h}^I=B^{1/2}h^I$ and we have used the identities
\begin{equation}
h^I_x = - \sqrt{\frac{3}{2\kappa^2_5}}h^I_{,x}, \quad h^Ih^x_I=0, \quad g_{xy}=a_{IJ}h^I_x h^J_y, \quad a_{IJ}=-2C_{IJK}+3h_Ih_J.
\end{equation}
In (\ref{orbifold3:eps}), $n_V+1$ vector multiplet scalars ${\phi}^x$ appear through $n_V+1$ scalars $\hat{h}^I$ subject
to the constraint,
\begin{equation}
C_{IJK}(\hat{h})\hat{h}^I\hat{h}^J\hat{h}^K=B^{-3/2}.
\end{equation}
\indent Now we  consider the low-energy effective action of
${\mathcal{N}}=1$ supergravity in four dimensions. Therefore, we
must show that the scalars in $(\ref{orbifold3:eps})$ parametrize
a complex manifold of the K\"{a}hler type. For that purpose, we
define two new complex quantities, $T$ and $S$:
\begin{eqnarray}\label{new:eps}
T^I&=& \frac{1}{\kappa_5} \hat{h}^I + i\sqrt{\frac{3}{2}}A^I_5, \\
S&=& \frac{1}{\kappa_5} \Big(\rho + i\eta \Big).\label{new1:eps}
\end{eqnarray}
From this definition, the function $f(\rho,\eta)$ is replaced by $f(S,\bar{S})$ and we get
\begin{eqnarray}\label{def:eps}
\hat{a}_{IJ}&=&\frac{6}{(T^I+\bar{T}^I)(T^J+\bar{T}^J)}, \\
f_{\rho}  &=& \frac{1}{\kappa_5} \Big( f_{S} + f_{\bar{S}} \Big), \\\label{def1:eps}
f_{\eta}  &=& \frac{i}{\kappa_5} \Big( f_{S} - f_{\bar{S}} \Big).\label{def2:eps}
\end{eqnarray}
The Laplace equation $(\ref{diff:eps})$ can be casted into
\begin{equation}
f_{S\bar{S}}=\frac{3f}{4(S+\bar{S})^2}.
\end{equation}
\indent By substituting $(\ref{new:eps})$ and $(\ref{def:eps})$
into $(\ref{orbifold3:eps})$, the action can be rewritten as
\begin{eqnarray}\label{orbifold4:eps}
S_{S^1/{\mathbb{Z}}_2}&=&
\int d^4x \sqrt{-g}\frac{1}{2\kappa^2_4}R
\nonumber\\
& &- \int d^4x \sqrt{-g} \frac{3}{\kappa^2_4 \big(T^I+ T^{\bar{I}}
\big)\big(T^J+ T^{\bar{J}}\big)}
\partial_{\mu}T^I \partial^{\mu} T^{\bar{J}}
\nonumber\\
& & - \int d^4x \sqrt{-g} \frac{1}{\kappa^2_4} \Bigg[
\frac{1}{(S+\bar{S})^2} + 2\Bigg(
\frac{f_{S}f_{\bar{S}}}{f^2}-\frac{f_{S\bar{S}}}{f} \Bigg) \Bigg]
\partial_{\mu}S \partial^{\mu}\bar{S} \nonumber\\
&=&
\frac{1}{2\kappa^2_4} \int d^4x \sqrt{-g} R
- \int d^4x \sqrt{-g} K_{I\bar{J}}\partial_{\mu}T^I \partial^{\mu} T^{\bar{J}} \nonumber\\
&& -\int d^4x \sqrt{-g} K_{S\bar{S}}\partial_{\mu}S \partial^{\mu}\bar{S},
\end{eqnarray}
where
\begin{eqnarray}
K_{I\bar{J}}&\equiv& \frac{\partial}{\partial T^I}\frac{\partial}{\partial T^{\bar{J}}}K_V, \\
K_{S\bar{S}}&\equiv& \frac{\partial}{\partial S}\frac{\partial}{\partial \bar{S}}K_H,
\end{eqnarray}
The K\"{a}hler potentials are denoted by $K_V$ and $K_H$ which
comes from the vector- and hypermultiplets, respectively. We find
that
\begin{eqnarray}\label{kler}
K &=& K_V + K_H \nonumber\\
&=&-\kappa^{-2}_4 ln \bigg( C_{IJK}(T^I+T^{\bar{I}})(T^J+T^{\bar{J}})(T^K+T^{\bar{K}})\bigg) \nonumber\\
& &-\kappa^{-2}_4 \bigg( ln(S+\bar{S})+2ln f \bigg).
\end{eqnarray}
\subsection{The fermionic sector}
In the previous subsection we found the effective action for the
bosonic sectors with the K\"{a}hler potentials is given by
$(\ref{kler})$. The rest is to derive the fermionic sectors of the
effective four dimensional ${\mathcal{N}}=1$ theory. Since the
ansatz metric is non-radion background, the fermionic fields do
not depend on $x^5$ and then the integral over the compact
dimensions in the action yields just the volume which can be
absorbed into the definition of the four dimensional gravitational
constant. In other words, it is equivalent to integrate out the
massive Kaluza-Klein modes and one keeps only zero modes in the effective description.\\
\indent Our starting point is the fermionic parts of the action
$(\ref{action})$. The kinetic term of the $\psi \psi$-component is
given by
\begin{equation}\label{gravitino}
S_{\psi\psi} =\int d^5x \sqrt{-\hat{g}} \Bigg[
-\frac{1}{2\kappa^2_5} \bar{\psi}_{\hat{\rho}}
\gamma^{\hat{\rho}\hat{\mu}\hat{\nu}} D_{\hat{\mu}}
\psi_{\hat{\nu}} - \frac{i\sqrt{6}}{16\kappa_5} h_{I}
\hat{F}^I_{\hat{\rho}\hat{\sigma}} \bar{\psi}_{\hat{\mu}}
\gamma^{\hat{\mu}\hat{\nu}\hat{\rho}\hat{\sigma}} \psi_{\hat{\nu}}
\Bigg].
\end{equation}
\indent It is convenient to combine two symplectic Majorana
spinors into one even(odd) Majorana spinor in four dimensions,
with the following convention
\begin{equation}
\psi_{\hat{\mu}}   =
\left(
\begin{array}{c}
\psi^1_{\hat{\mu} L}
\\
\phantom{bla}
\\
\overline{\psi}^{2 }_{\hat{\mu} R}
\end{array}
\right), \qquad
\overline{\psi}_{\hat{\mu}}  =
\left(
\psi^{2 }_{\hat{\mu} L},
\overline{\psi}^1_{\hat{\mu} R}
\right).
\end{equation}\label{spinor5}
Using these definitions, we can rewrite ($\ref{gravitino}$) in
terms of even and odd combinations,
\begin{equation}
S_{\psi\psi} =\int d^5x \sqrt{-\hat{g}} \Bigg[
-\frac{1}{2\kappa^2_5} \psi^2_{\hat{\rho}}
\gamma^{\hat{\rho}\hat{\mu}\hat{\nu}} D_{\hat{\mu}}
\psi^1_{\hat{\nu}} - \frac{i\sqrt{6}}{16\kappa_5} h_{I}
\hat{F}^I_{\hat{\rho}\hat{\sigma}} \psi^2_{\hat{\mu}}
\gamma^{\hat{\mu}\hat{\nu}\hat{\rho}\hat{\sigma}}
\psi^1_{\hat{\nu}} \Bigg] + h.c.,
\end{equation}
or
\begin{eqnarray}\label{gravitino1}
S_{\psi\psi}  &=& \int d^5x \sqrt{-\hat{g}}\Bigg[
-\frac{1}{2\kappa^2_5} .\frac{1}{2}. \left( \psi^{+}_{\hat{\rho}}
\gamma^{\hat{\rho}\hat{\mu}\hat{\nu}} D_{\hat{\mu}}
\psi^{+}_{\hat{\nu}}
- \psi^{-}_{\hat{\rho}} \gamma^{\hat{\rho}\hat{\mu}\hat{\nu}} D_{\hat{\mu}} \psi^{-}_{\hat{\nu}} \right) \nonumber\\
&&- \frac{i\sqrt{6}}{16\kappa_5} h_{I}
\hat{F}^I_{\hat{\rho}\hat{\sigma}} \frac{1}{2} \left(
\psi^{+}_{\hat{\mu}}
\gamma^{\hat{\mu}\hat{\nu}\hat{\rho}\hat{\sigma}}
\psi^{+}_{\hat{\nu}} - \psi^{-}_{\hat{\mu}}
\gamma^{\hat{\mu}\hat{\nu}\hat{\rho}\hat{\sigma}}
\psi^{-}_{\hat{\nu}} \right) \Bigg] + h.c.,
\end{eqnarray}
where we have used
\begin{equation}
\psi^{1}_{\hat{\mu}}=\frac{1}{\sqrt{2}}\left(\psi^{+}_{\hat{\mu}} + \psi^{-}_{\hat{\mu}} \right) \qquad
\psi^{2}_{\hat{\mu}}=\frac{1}{\sqrt{2}}\left(\psi^{+}_{\hat{\mu}} - \psi^{-}_{\hat{\mu}} \right).
\end{equation}
The two fixed points require that certain ${\mathbb{Z}}_2$-odd
fields have a step-function, namely  $sgn(z=x^5)$.
The step function $sgn(z=x^5)$ takes values $-1$ for $x^5 \in [-\pi R, 0]$ and $+1$ for $x^5 \in [0, \pi R]$.\\
\indent In order for the reduction to be consistent, we must make
an ansatz for the gravitino as follows
\begin{equation}
\psi^{+}_{\mu}=\frac{1}{\sqrt{2}} C(x) \psi_{\mu}, \qquad
\psi^{-}_{\mu}=\frac{1}{\sqrt{2}} sgn(z) C(x) \psi_{\mu}.
\end{equation}
where $C(x)$ is an arbitrary function. After subtracting to each
component and plugging the ansatz into ($\ref{gravitino1}$) we
get\footnote{The detailed calculation is presented in Appendix C.}
\begin{eqnarray}\label{gravitino2}
S_{\psi\psi}  &=& \int d^5x \sqrt{-\hat{g}} \Bigg[
-\frac{1}{2\kappa^2_5} \frac{1}{2} \left( \psi^{+}_{\rho}
\gamma^{\rho\mu\nu} D_{\mu} \psi^{+}_{\nu} + \psi^{+}_{\rho}
\gamma^{\rho 5 \nu} D_{5} \psi^{-}_{\nu} \right)
- \frac{i\sqrt{6}}{16\kappa_5} h_{I} \hat{F}^I_{\hat{\rho}\hat{\sigma}} \frac{1}{2}  \psi^{+}_{\mu} \gamma^{\mu\nu \hat{\rho}\hat{\sigma}} \psi^{+}_{\nu} \nonumber\\
&&+\frac{1}{2\kappa^2_5} \psi^{+}_{\rho} \gamma^{\rho\nu}
\psi^{+}_{\nu} \left( \delta(z) - \delta(z-\pi R) \right) \Bigg] +
h.c..
\end{eqnarray}
We note that the third term of the action ($\ref{gravitino2}$) is
the boundary term. This result has to be consistent with the
upstair approach used in \cite{HW} where one regards space-time as
the smooth manifold $M_4 \times S^1$ subject to ${\mathbb{Z}}_2$
symmetry. Then, in the framework of the five dimensional
supergravity we can write the total action by
\begin{equation}
S = S_{bulk} + S_{boundary},
\end{equation}
where $S_{bulk}$ is given by ($\ref{action}$) and $S_{boundary}$
or $S_{brane}$ in the context of braneworld is given by
\begin{equation}
S_{brane} = \frac{1}{2\kappa^2_5} \int d^5x \sqrt{-g} \psi^{+}_{\mu} \gamma^{\mu\nu} \psi^{+}_{\nu} [ \delta(z) - \delta(z - \pi R) ] + h.c.
\end{equation}
\indent The bulk part of the action ($\ref{gravitino2}$) can be
rewritten as follows
\begin{equation}\label{psi1}
S_{\psi\psi}  = \frac{1}{2\kappa^2_5} \int d^5x \sqrt{-\hat{g}}
 \frac{C^2}{4} \psi_{\rho} \gamma^{\rho\mu\nu} \tilde{D}_{\mu} \psi_{\nu} + h.c.,
\end{equation}
where
\begin{eqnarray}\label{covder1}
\tilde{D}_\mu \psi_{\nu} &=& \partial_{\mu}\psi_{\nu} + \frac{1}{2}\omega_{\mu}{}^{ab} \gamma_{ab} \psi_{\nu} + \partial_{\mu} ln C \psi_{\nu}
- \frac{1}{4} \partial_{\mu} ln B \psi_{\nu} \nonumber\\
&&- \frac{\kappa^2_4}{12} ( K_I \partial_{\mu} T^I - K_{\bar{I}} \partial_{\mu} T^{\bar{I}}) \psi_{\nu}
+ \frac{i\kappa^2_4}{2} ( K_{S} \partial_{\mu} S - K_{\bar{S}} \partial_{\mu} \bar{S} ) \psi_{\nu}.
\end{eqnarray}
After integrating (\ref{psi1}) with respect to $x^5 \in [-\pi
R,\pi R]$, we finally get
\begin{equation}
S_{\psi\psi}  = -\frac{1}{2\kappa^2_4} \int d^4x \sqrt{-g}
\overline{\psi}_{\rho} \gamma^{\rho\mu\nu} \tilde{D}_{\mu} \psi_{\nu}.
\end{equation}
In the above expressions we have set $C = B^{1/4}$ so that the
covariant derivative in four dimensions becomes
\begin{eqnarray}\label{covder2}
\tilde{D}_\mu \psi_{\nu} &=& \partial_{\mu}\psi_{\nu} +
\frac{1}{2}\omega_{\mu}{}^{ab} \gamma_{ab} \psi_{\nu} -
\frac{\kappa^2_4}{12} ( K_I \partial_{\mu} T^I - K_{\bar{I}}
\partial_{\mu} T^{\bar{I}}) \psi_{\nu} \nonumber\\
&&+ \frac{i\kappa^2_4}{2} ( K_{S} \partial_{\mu} S - K_{\bar{S}}
\partial_{\mu} \bar{S} ) \psi_{\nu},
\end{eqnarray}
and then used the spin connections in the metric background,
\begin{eqnarray}
\hat{\omega}_{\mu}{}^{ab} &=& \omega_{\mu}{}^{ab} - \frac{1}{4} (e^a_{\mu} e^{\nu b} - e^b_{\mu} e^{\nu a}) \partial_{\nu} ln B, \\
\hat{\omega}_{5a\bar{5}} &=& - \frac{1}{2} B^{1/2} e^{\mu}_a \partial_{\mu} ln B,
\end{eqnarray}
and $Sp(1)$-connections,
\begin{equation}
\omega^1 = -\frac{f_{\eta}}{f} d\rho + \Big( \frac{1}{2\rho} + \frac{f_{\rho}}{f} \Big) d\eta.
\end{equation}
\indent The hyperino kinetic terms of the fermionic sector of the
action ($\ref{action}$) is given by
\begin{equation}\label{gaugino}
S_{\zeta\zeta}= \int d^5x \sqrt{-\hat{g}} \Bigg[- \bar{\zeta}^A
\gamma^{\hat{\mu}} D_{\hat{\mu}} \zeta_A +
\frac{i\kappa_5}{8}\sqrt{6}h_I\bar{\zeta}_A\gamma^{\hat{a}\hat{b}}
\hat{F}^I_{\hat{a}\hat{b}}\zeta^A \Bigg], \qquad A=1,2.
\end{equation}
We assume that the fields are independent of the fifth coordinate
so that derivative with respect to fifth coordinate vanishes. The
ansatz for the hyperino (even fields),
\begin{equation}
\zeta_A \rightarrow \zeta_1= \frac{1}{\sqrt{2}\kappa_5} B^{1/4} \zeta,
\end{equation}
and we obtain
\begin{equation}\label{gaugino1}
S_{\zeta\zeta}= - \frac{1}{2\kappa^2_4} \int d^4 x \sqrt{-g}
\bar{\zeta} \gamma^{\mu} \tilde{D}_{\mu} \zeta.
\end{equation}
The covariant derivative for the hyperino in four dimensions is
given by
\begin{equation}\label{covder3}
\tilde{D}_{\mu} \zeta = \partial_{\mu} \zeta + \frac{1}{2} \omega_{\mu}{}^{ab} \gamma_{ab} \zeta
- \frac{i\kappa^2_4}{2}(K_{S} \partial_{\mu}S - K_{\bar{S}} \partial_{\mu}{\bar{S}})\zeta
+ \frac{\kappa^2_4}{6}(K_{I} \partial_{\mu} T^I - K_{\bar{I}} \partial_{\mu} T^{\bar{I}}) \zeta.
\end{equation}
Next we look at the gaugino kinetic terms
\begin{equation}
S_{\lambda\lambda}= \int d^5x \sqrt{-\hat{g}} \Bigg[ - \frac{1}{2}
\bar{\lambda}_x \gamma^{\hat{\mu}} D_{\hat{\mu}} \lambda^x +
\frac{i\kappa_5}{4} \sqrt{\frac{2}{3}} \Bigg(\frac{1}{4} g_{xy}
h_I + T_{xyz} h^z_I \Bigg) \lambda^x \gamma^{\hat{a}\hat{b}}
\hat{F}^{I}_{\hat{a}\hat{b}} \lambda^y \Bigg].
\end{equation}
For gaugino, the procedure is similar to the hyperino terms and we
then obtain
\begin{equation}
S_{\lambda\lambda}= - \frac{1}{2\kappa^2_4} \int d^4x \sqrt{-g} \bar{\lambda^x} \gamma^{\mu} \tilde{D}_{\mu} \lambda_x,
\end{equation}
where
\begin{eqnarray}\label{covder4}
\tilde{D}_{\mu} \lambda_x &=& \partial_{\mu} \lambda_x +  \frac{1}{2} \omega_{\mu}{}^{ab} \gamma_{ab} \lambda_x +
 \partial_{\mu} \phi^y \Gamma^z_{yx}  \lambda_z
-\frac{i\kappa^2_4}{2} (K_{S} \partial_{\mu} S - K_{\bar{S}} \partial_{\mu} \bar{S}) \lambda_x \nonumber\\
&&+ \frac{\kappa^2_4}{12} (K_{I}\partial_{\mu}T^I - K_{\bar{I}} \partial_{\mu} T^{\bar{I}}) \lambda_x.
\end{eqnarray}
Finally, by combining the results of the bosonic and the fermionic
sectors we get the low-energy effective action of
${\mathcal{N}}=1$ in four dimensions:
\begin{eqnarray}\label{orbifold5:eps}
S^{N = 1}_{d=4} &=& \frac{1}{2\kappa^2_4} \int d^4x \sqrt{-g}
\Bigg[R + \overline{\psi}_{\rho} \gamma^{\rho\mu\nu}
\tilde{D}_{\mu}\psi_{\nu} - \bar{\zeta} \gamma^{\mu}
\tilde{D}_{\mu} \zeta
-\bar{\lambda^x} \gamma^{\mu} \tilde{D}_{\mu} \lambda_x \nonumber\\
&&- 2 \kappa^2_4 K_{I\bar{I}}\partial_{\mu} T^I \partial^{\mu}
T^{\bar{I}} - 2 \kappa^2_4 K_{S\bar{S}}\partial_{\mu} S
\partial^{\mu} \bar{S} \Bigg],
\end{eqnarray}
where the covariant derivative of the spinors are given by ($\ref{covder2}$), ($\ref{covder3}$) and ($\ref{covder4}$).
\section{Conclusions and Outlook}
In this paper we have studied $S^1/{\mathbb{Z}}_2$
compactification of the ungauged ${\mathcal{N}}=1$ supergravity in
five dimensions coupled to arbitrary vector multiplets and a
hypermultiplet where the scalar fields span toric self dual
Eintein spaces. The resulting theory is four dimensional
${\mathcal{N}}=1$ supergravity. In the bosonic sector we found
that  K\"{a}hler potential from the vector multiplets contribution
has the form
\begin{equation}
K_V = -\kappa^{-2}_4 ln \left ( C_{IJK}(T^I+T^{\bar{I}})(T^J+T^{\bar{J}})(T^K+T^{\bar{K}})  \right),
\end{equation}
and from hypermultiplets contribution are
\begin{equation}
K_H= -\kappa^{-2}_4 \left( ln(S+\bar{S})+2ln f \right).
\end{equation}
\indent Furthermore, we have derived the fermionic sectors of the
four dimensional ${\mathcal{N}}=1$ theory. In general, our results
confirm those in reference \cite{Bagg1}, where the
effective theory was obtained by compactification from five dimensional supergravity down to four dimensions in the context of Randall-Sundrum scenario.  \\
\indent It is also interesting to extend our scenario to the
general case where the ${\mathcal{N}}=2$ scalar potential reduced
to the ${\mathcal{N}}=1$ scalar potential in terms of holomorphic
superpotential. Another problem is to cancel anomaly of the
orbifold model and find the gauge group on two orbifold fixed
points such that the gauge and the scalar fields residing on the
boundaries can be supersymmetrized. This can be done by modifying
the boundary action but also one has to modify the supersymmetry
transformation
laws in both boundary and bulk fields. This problem will be addressed elsewhere.\\
\indent Note Added: During preparation of this manuscript we became aware of the independent work \cite{BN04} which has some overlaps with our results.\\

\newpage
\appendix
\section{Conventions and Notations}
The purpose of this appendix is to collect  our conventions in this paper. The
spacetime metric is taken to have the signature $(-,+,+,+,+)$ while the Ricci
scalar is defined to be
$R=g^{\mu\nu}\Big[\frac{1}{2}\partial_{\rho}\Gamma^{\rho}_{\mu\nu}
-\partial_{\nu}\Gamma^{\rho}_{\mu\rho}+\Gamma^{\sigma}_{\mu\nu}\Gamma^{\rho}_{\sigma\rho}\Big]
+\frac{1}{2}\partial_{\rho}\Big[g^{\mu\nu}\Gamma^{\rho}_{\mu\nu}\Big]$.
The Christoffel symbol is given by
$\Gamma^{\mu}_{\nu\rho}=\frac{1}{2}g^{\mu\sigma}(\partial_{\nu}g_{\rho\sigma}+\partial_{\rho}g_{\nu\sigma}-\partial_{\sigma}g_{\nu\rho})$
where $g_{\mu\nu}$ is the spacetime metric.\\
\indent \textit{Indices}\\
\begin{tabular}{r @{\hspace{2.5 cm}}  l }
$\hat{\mu},\hat{\nu} = 0,...,3,5$ & label curved five dimensional spacetime indices \\
\\
$\hat{a}, \hat{b} = 0,...,3,5$ & label flat five dimensional spacetime indices \\
\\
$\mu, \nu = 0,...,3$ & label curved four dimensional spacetime indices \\
\\
$a, b = 0,...,3$ & label flat four dimensional spacetime indices \\
\\
$i,j=1,2$ & label the fundamental representation of \\
&  the $R$-symmetry group $SU(2) \otimes U(1)$ \\
\\
$x,y,z=1,..,n_V$ & label the real scalars of the ${\mathcal{N}}=2$ vector multiplet \\
\\
$I,J,K=0,1,..,n_V$ & label the  vector multiplets  \\
\\
$X,Y,Z =1,...,4n_H$ & label the real scalars of the ${\mathcal{N}}=2$ hypermultiplet\\
\\
$A, B=1,...,2n_H$ & label the fundamental representation of $Sp(2n_H)$\\
\\
\end{tabular}
\section{Five dimensional Supergravity on $S^1$}
In this section, we derive the dimensional reduction of the
bosonic part of the five dimensional supergravity action on
$S^1$.\footnote{This  has also been  discussed in the appendix of
\cite{AFL04}.} The class of four dimensional theories obtained in
this way are only a subclass of the general four dimensional
${\mathcal{N}}=2$ theories \cite{ABCD}.
In general, we can choose the metric to be
\begin{eqnarray}\label{eqansatz:eps}
ds^2_5&=&A(x)g_{\mu\nu}dx^{\mu}dx^{\nu}+B(x)dz^2 \nonumber\\
&=&\hat{g}_{\hat{\mu}\hat{\nu}}dx^{\hat{\mu}}dx^{\hat{\nu}}
\end{eqnarray}
where $A$ and $B$ are arbitrary functions. We have
\begin{equation}
\hat{g}_{\mu\nu}=A(x)g_{\mu\nu} \quad \hat{g}_{zz}=B(x).
\end{equation}
The gravity term in the supergravity Lagrangian reduced to
\begin{equation}
\sqrt{-\hat{g}}\hat{R} \sim
\sqrt{-g}\bigg[AB^{1/2}R+\frac{3}{2}A^{-1}B^{1/2}\partial_{\mu}{A}\partial^{\mu}{A}
+\frac{3}{2}B^{-1/2}\partial_{\mu}{A}\partial^{\mu}{B}\bigg],
\end{equation}
where $\sim$ means equal up to a total derivative.
The FF-term we are going to reduce is
\begin{displaymath}
{\mathcal{L}}_{FF}=-\frac{1}{4}\sqrt{-\hat{g}}a_{IJ}\hat{F}^I_{\hat{\mu}\hat{\nu}}\hat{F}^{\hat{\mu}\hat{\nu}J}.
\end{displaymath}
In order to get the massless sector we divide
$\hat{F}^I_{\hat{\mu}\hat{\nu}}$ into
$(\hat{F}^I_{\mu\nu},\hat{F}^I_{\mu 5})$. Then, we take Fourier
expansion of this term and keep only the lowest order terms which
should be independent of $x^5$. We then have that
\begin{eqnarray}
-\frac{1}{4}\sqrt{-\hat{g}}a_{IJ}\hat{F}^I_{\hat{\mu}\hat{\nu}}\hat{F}^{\hat{\mu}\hat{\nu}J}
&=&-\frac{1}{4}\sqrt{-g}B^{1/2}a_{IJ}F^I_{\mu\nu}F^{\mu\nu J} \nonumber\\
& &-\frac{1}{2}\sqrt{-g}AB^{-1/2}a_{IJ}\partial_{\mu}A^I_5\partial^{\mu}A^J_5,
\end{eqnarray}
where $A^I_5$ are the fifth component of the vector fields.
The vector multiplet scalars sector  reduce to
\begin{eqnarray}
{\mathcal{L}}_\phi&=&-\frac{1}{2}\sqrt{-\hat{g}}g_{xy}\partial_{\hat{\mu}}\phi^x\partial^{\hat{\mu}}\phi^y\nonumber\\
&=&-\frac{1}{2}\sqrt{-g}AB^{1/2}g_{xy}\partial_{\mu}\phi^x\partial^{\mu}\phi^y .
\end{eqnarray}
The hypermultiplet sector  reduce to
\begin{eqnarray}
{\mathcal{L}}_q&=&-\frac{1}{2}\sqrt{-\hat{g}}g_{XY}\partial_{\hat{\mu}}q^X\partial^{\hat{\mu}}q^Y\nonumber\\
&=&-\frac{1}{2}\sqrt{-g}AB^{1/2}g_{XY}\partial_{\mu}q^X\partial^{\mu}q^Y .
\end{eqnarray}
Moreover The Chern-Simon term can be simplified
\begin{eqnarray}
{\mathcal{L}}_{FFA}&=& \frac{1}{6\sqrt{6}}\epsilon^{\hat{\mu}\hat{\nu}\hat{\rho}\hat{\sigma}\hat{\lambda}}
C_{IJK}\hat{F}^I_{\hat{\mu}\hat{\nu}}\hat{F}^J_{\hat{\rho}\hat{\sigma}}A^K_{\hat{\lambda}} \nonumber\\
&=&\frac{1}{2\sqrt{6}}{\epsilon}^{\mu\nu\rho\sigma}
C_{IJK}\hat{F}^I_{\mu\nu}\hat{F}^J_{\rho\sigma}A^K_5 \nonumber\\
&=&\frac{1}{2\sqrt{6}}\epsilon^{\mu\nu\rho\sigma}
C_{IJK}F^I_{\mu\nu}F^J_{\rho\sigma}A^K_5.
\end{eqnarray}
Collecting the above results, the bosonic part of the reduced
supergravity Lagrangian $(\ref{action})$ is
\begin{eqnarray}\label{eqs1:eps}
\frac{1}{\sqrt{-g}}{\mathcal{L}}^{S^1}_{4d}&=&
-\frac{1}{2}AB^{1/2}R-\frac{3}{4}\partial_\mu ln A\partial^\mu ln A
-\frac{3}{4}\partial_\mu ln A\partial^\mu ln B \nonumber\\
& &-\frac{1}{2}AB^{1/2}g_{xy}\partial_\mu\phi^x\partial^\mu\phi^y
-\frac{1}{4}B^{1/2}a_{IJ}F^I_{\mu\nu}F^{\mu\nu J} \nonumber\\
& &-\frac{1}{2}AB^{-1/2}a_{IJ}\partial_{\mu}A^I_5\partial^{\mu}A^J_5
-\frac{1}{\sqrt{6}}C_{IJK}\tilde{F}^I_{\mu\nu}F^{\mu\nu J}A^K_5.
\end{eqnarray}
Here, we only consider the arbitrary function $A=A(x)$ and $B=B(x)$.
To put the equation (\ref{eqs1:eps}) back into the canonical form,
we perform a Weyl rescaling of the metric which is given by:
\begin{equation}
g_{\mu\nu}\rightarrow A^{-1}B^{-1/2}g_{\mu\nu}.
\end{equation}
So under this rescaling the lagrangian becomes:
\begin{eqnarray}\label{eqs2:eps}
\frac{1}{\sqrt{-g}}{\mathcal{L}}^{S^1}_{4d}&=&\frac{1}{2}R-\hat{a}_{IJ}(\hat{h^I})\partial_\mu\hat{h}^I\partial^\mu\hat{h}^J
-\frac{2}{3}\hat{a}_{IJ}(\hat{h^I})\partial_\mu A^I_5 \partial^\mu A^J_5 \nonumber\\
& &-\frac{1}{4}B^{1/2}a_{IJ}F_{\mu\nu}^IF^{\mu\nu{J}}
+\frac{1}{\sqrt{6}}C_{IJK}\tilde{F}^I_{\mu\nu}F^{\mu\nu J}A_5^K,
\end{eqnarray}
where $\hat{a}_{IJ}=\frac{3}{4}B^{-1}a_{IJ}$ and
$\hat{h}^I=B^{1/2}h^I$. The Chern-Simons term vanishes identically
for the non abelian case.

\section{Some detailed calculations of the fermionic sectors}
The five dimensional spinor $\Psi$ and its conjugate
$\overline{\Psi}$ can be written into four dimensional form which
are $\Psi \equiv (\psi^1_{L}, {\overline{\psi^2}}_{R})^T$ and
$\overline{\Psi} \equiv (\psi^{2 L}, {\overline{\psi^1}}_{R})$.
Plugging this definition into ($\ref{gravitino1}$) we get
\begin{eqnarray}\label{c1}
S_{\psi\psi} &=& \int d^5x \sqrt{-\hat{g}} \Bigg[
-\frac{1}{2\kappa^2_5} \psi^{2}_{\hat{\rho}}
\gamma^{\hat{\rho}\hat{\mu}\hat{\nu}}
D_{\hat{\mu}}\psi^1_{\hat{\nu}}
- \frac{1}{2\kappa^2_5} \overline{\psi}^{1}_{\hat{\rho}} \gamma^{\hat{\rho}\hat{\mu}\hat{\nu}} D_{\hat{\mu}} \overline{\psi}^2_{\hat{\nu}}\nonumber\\
&&- \frac{i\sqrt{6}}{16\kappa_5} h_{I} \hat{F}^I_{\hat{\rho}\hat{\sigma}} \psi^2_{\hat{\mu}}
 \gamma^{\hat{\mu}\hat{\nu}\hat{\rho}\hat{\sigma}} \psi^1_{\hat{\nu}}
-\frac{i\sqrt{6}}{16\kappa_5} h_{I} \hat{F}^I_{\hat{\rho}\hat{\sigma}} \overline{\psi}^1_{\hat{\mu}}
 \gamma^{\hat{\mu}\hat{\nu}\hat{\rho}\hat{\sigma}} \overline{\psi}^2_{\hat{\nu}}
\Bigg].
\end{eqnarray}
By using $\psi^1_{\hat{\mu}} = \frac{1}{\sqrt{2}} (
\psi^{+}_{\hat{\mu}} + \psi^{-}_{\hat{\mu}})$ and
$\psi^2_{\hat{\mu}} = \frac{1}{\sqrt{2}} ( \psi^{+}_{\hat{\mu}} -
\psi^{-}_{\hat{\mu}})$, (\ref{c1}) can be rewritten as
\begin{eqnarray}\label{c2}
S_{\psi\psi} &=&\int d^5x \sqrt{-\hat{g}} \Bigg[
-\frac{1}{2\kappa^2_5} \frac{1}{2} \Bigg( \psi^{+}_{\hat{\rho}}
\gamma^{\hat{\rho}\hat{\mu}\hat{\nu}} D_{\hat{\mu}}
\psi^{+}_{\hat{\nu}} - \bar{\psi}^{-}_{\hat{\rho}}
 \gamma^{\hat{\rho}\hat{\mu}\hat{\nu}} D_{\hat{\mu}} \psi^{-}_{\hat{\nu}} \Bigg)\nonumber\\
&&- \frac{i\sqrt{6}}{16\kappa_5} h_{I}
\hat{F}^I_{\hat{\rho}\hat{\sigma}} \frac{1}{2} \Bigg(
\psi^{+}_{\hat{\mu}}
\gamma^{\hat{\mu}\hat{\nu}\hat{\rho}\hat{\sigma}}
\psi^{+}_{\hat{\nu}} - \psi^{-}_{\hat{\mu}}
\gamma^{\hat{\mu}\hat{\nu}\hat{\rho}\hat{\sigma}}
\psi^{-}_{\hat{\nu}} \Bigg) \Bigg] + h.c..
\end{eqnarray}
Now we split the space-time coordinates into: $x^{\hat{\mu}}=(x^\mu,x^5)$,
\begin{eqnarray}\label{c3}
S_{\psi\psi} &=& \int d^5x \sqrt{-\hat{g}} \Bigg[
-\frac{1}{2\kappa^2_5} \frac{1}{2}  \Bigg( \psi^{+}_{\rho}
\gamma^{\rho\mu\nu} D_{\mu} \psi^{+}_{\nu} + \psi^{+}_{\rho}
\gamma^{\rho\mu 5} D_{\mu} \psi^{+}_{5} \nonumber\\
&&+ \psi^{+}_{\rho} \gamma^{\rho 5 \nu} D_5 \psi^{+}_{\nu} +
\psi^{+}_{\rho} \gamma^{\rho 55} D_{5} \psi^{+}_{5}  +
\psi^{+}_{5} \gamma^{5 \mu\nu} D_{\mu} \psi^{+}_{\nu} +
\psi^{+}_{5} \gamma^{5 5 \nu} D_{5} \psi^{+}_{\nu} \nonumber\\
&&+ \psi^{+}_{5} \gamma^{5 \mu 5} D_{\mu} \psi^{+}_{5}
+\psi^{+}_{5}\gamma^{5 5 5} D_{5} \psi^{+}_{5} \Bigg)
-\frac{i\sqrt{6}}{16\kappa_5}h_{I}\hat{F}^I_{\hat{\rho}\hat{\sigma}}
\frac{1}{2}\Bigg( \psi^{+}_{\mu}
\gamma^{\mu\nu\hat{\rho}\hat{\sigma}} \psi^{+}_{\nu}
\nonumber\\
&&+\psi^{+}_{\mu} \gamma^{\mu 5 \hat{\rho}\hat{\sigma}}
\psi^{+}_{5} +\psi^{+}_{5} \gamma^{5 \nu\hat{\rho}\hat{\sigma}}
\psi^{+}_{\nu} +\psi^{+}_{5} \gamma^{5 5 \hat{\rho}\hat{\sigma}}
\psi^{+}_{5} \Bigg) - (+ \rightarrow-)\Bigg] + h.c..
\end{eqnarray}
By inserting the ansatz for the gravitino $\psi^{-}_{\mu} = sgn(z)
\psi^{+}_{\mu}$ and $\psi^{-}_5 = sgn(z) \psi^{+}_5$, (\ref{c3})
becomes
\begin{eqnarray}\label{c4}
S_{\psi\psi} &=& \int d^5x \sqrt{-\hat{g}} \Bigg[
-\frac{1}{2\kappa^2_5} \frac{1}{2}  (1 - sgn(z)^2) \Bigg(
\psi^{+}_{\rho} \gamma^{\rho\mu\nu} D_{\mu} \psi^{+}_{\nu} +
\psi^{+}_{\rho} \gamma^{\rho 5 \nu} D_5 \psi^{+}_{\nu}
\Bigg)\nonumber\\
&& + \frac{1}{2\kappa^2_5} sgn(z) \delta(z) \psi^{+}_{\rho}
\gamma^{\rho\nu} \psi^{+}_{\nu} - \frac{i\sqrt{6}}{16\kappa_5}
h_{I} \hat{F}^I_{\hat{\rho}\hat{\sigma}} \frac{1}{2} (1 -
sgn(z)^2) \psi^{+}_{\mu} \gamma^{\mu\nu\hat{\rho}\hat{\sigma}}
\psi^{+}_{\nu} \Bigg] \nonumber\\ && + h.c.,
\end{eqnarray}
where we have used $\partial_5 sgn(z)=2\delta(z)- \delta(z - \pi R)$. For any values of the $sgn(z)$ we get
\begin{eqnarray}\label{c5}
S_{\psi\psi} &=& \int d^5x \sqrt{-\hat{g}} \Bigg[
-\frac{1}{2\kappa^2_5} \frac{1}{2} \Bigg( \psi^{+}_{\rho}
\gamma^{\rho\mu\nu} D_{\mu} \psi^{+}_{\nu} + \psi^{+}_{\rho}
\gamma^{\rho 5 \nu} D_5 \psi^{+}_{\nu}
\Bigg)\nonumber\\
&& - \frac{i\sqrt{6}}{16\kappa_5} h_{I}
\hat{F}^I_{\hat{\rho}\hat{\sigma}} \frac{1}{2} \psi^{+}_{\mu}
\gamma^{\mu\nu\hat{\rho}\hat{\sigma}} \psi^{+}_{\nu} -
\frac{1}{2\kappa^2_5}\psi^{+}_{\rho} \gamma^{\rho\nu}
\psi^{+}_{\nu} [\delta(z) - \delta(z - \pi R)]
\Bigg] \nonumber\\
&& + h.c..
\end{eqnarray}
Next, compactification of the bulk gravitino can be done by
inserting the ansatz $\psi^{+}_\mu = \frac{1}{\sqrt{2}} C(x)
\psi_\mu$ into (\ref{c5}) and by using the covariant derivative of
the gravitino (\ref{cov3}), one obtains
\begin{eqnarray}\label{c6}
S_{\psi\psi} &=& - \frac{1}{2\kappa^2_5} \int d^5x \sqrt{-\hat{g}}
\frac{C^2}{4} \psi_{\rho} \gamma^{\rho\mu\nu} \Bigg(
\partial_{\mu} \psi_{\nu} + \frac{1}{2}
\omega_{\mu}{}^{ab} \gamma_{ab}\psi_{\nu} \Bigg) \nonumber\\
&&+\frac{1}{2\kappa^2_5} \int d^5x \sqrt{-\hat{g}} \frac{C^2}{4}
\psi_{\rho} \gamma^{\rho\mu\nu} \Bigg( -\frac{f_\eta}{f} \partial_{\mu} \rho + \Bigg( \frac{1}{2 \rho} + \frac{f_\rho}{f} \Bigg)\partial_{\eta} \Bigg) \psi_{\nu}\nonumber\\
&&- \frac{1}{2\kappa^2_5} \int d^5x \sqrt{-\hat{g}} \frac{C^2}{4}
\psi_{\rho} \gamma^{\rho\mu\nu} \Bigg( \partial_{\mu} ln C - \frac{1}{4} \partial_{\mu} ln B \Bigg)\nonumber\\
&&- \frac{1}{2\kappa^2_5} \int d^5x \sqrt{-\hat{g}} \frac{C^2}{4}
\frac{i \sqrt{6}}{4 \kappa_5} h_I \partial_{\rho} A^I_5 \psi_{\mu}
\gamma^{\mu\nu\rho} \psi_{\nu} + h.c..
\end{eqnarray}
From the bosonic sectors we have obtained the equation
(\ref{new:eps}) - (\ref{def2:eps}). Inserting these equations into
(\ref{c6}), one obtains
\begin{eqnarray}\label{c7}
S_{\psi\psi} &=& - \frac{1}{2\kappa^2_5} \int d^5x \sqrt{-\hat{g}}
\frac{C^2}{4} \psi_{\rho} \gamma^{\rho\mu\nu} \Bigg[
\partial_{\mu} \psi_{\nu} + \frac{1}{2} \omega_{\mu}{}^{ab}
\gamma_{ab}\psi_{\nu}
 + \partial_{\mu} ln C \psi_{\nu} - \frac{1}{4} \partial_{\mu} ln B \psi_{\nu} \nonumber\\
&& - \frac{i}{2} \Bigg( \Bigg( \frac{2 f_S}{f} + \frac{1}{S +
\bar{S}} \Bigg) \partial_{\mu} S
- \Bigg( \frac{2 f_{\bar{S}}}{f} + \frac{1}{S + \bar{S}} \Bigg) \partial_{\mu} \bar{S} \Bigg) \psi_{\nu} \nonumber\\
&& + \Bigg( \frac{1}{4(T^I + T^{\bar{I}})} \partial_{\mu} T^I -
\frac{1}{4(T^I + T^{\bar{I}})} \partial_{\mu} T^{\bar{I}} \Bigg)
\psi_{\nu} \Bigg] + h.c.,
\end{eqnarray}
where the K\"{a}hler potential (\ref{kler}) we have
\begin{eqnarray}
K_S &\equiv& \frac{\partial K}{\partial S}=- \frac{1}{\kappa^2_4}
\left( \frac{2 f_S}{f} + \frac{1}{S + \bar{S}} \right),\\
K_{\bar{S}} &\equiv&  \frac{\partial K}{\partial {\bar{S}}} =- \frac{1}{\kappa^2_4} \left( \frac{2f_{\bar{S}}}{f} + \frac{1}{S + {\bar{S}}} \right),\\
K_I &\equiv&  \frac{\partial K}{\partial T^I}=-
\frac{3}{\kappa^2_4}
\left(  \frac{1}{T^I + T^{\bar{I}}} \right),\\
K_{\bar{I}} &\equiv& \frac{\partial K}{\partial T^{\bar{I}}} =-
\frac{3}{\kappa^2_4} \left( \frac{1}{T^I+ T^{\bar{I}}} \right),
\end{eqnarray}
so that we get the equation (\ref{psi1}). The spinors in four
dimensions are then defined by
\begin{equation}
\psi_{\mu}   = \frac{1}{2}
\left(
\begin{array}{c}
\psi_{\mu L}
\\
\phantom{bla}
\\
\overline{\psi}_{\mu R}
\end{array}
\right), \qquad
\overline{\psi}_{\mu}  = \frac{1}{2}
\left(
\psi_{\mu L},
\overline{\psi}_{\mu R}
\right),
\end{equation}\label{spinor4}
and one obtains
\begin{equation}\label{c8}
S_{\psi\psi} = - \frac{1}{2\kappa^2_5} \int d^5x \sqrt{-\hat{g}} C^2
\bar{\psi}_{\rho} \gamma^{\rho\mu\nu} \tilde{D}_{\mu} \psi_{\nu} .
\end{equation}
Integrating out the above equation with respect to $x^5 \in [-\pi
R,\pi R]$, it finally results
\begin{eqnarray}
S_{\psi\psi} &=& - \frac{1}{2\kappa^2_5} \int d^4x \int_0^{2\pi R} dz \sqrt{-\hat{g}} C^2
\bar{\psi}_{\rho} \gamma^{\rho\mu\nu} \tilde{D}_{\mu} \psi_{\nu} \nonumber\\
&=& - \frac{1}{2\kappa^2_4} \int d^4x  \sqrt{-g}
\bar{\psi}_{\rho} \gamma^{\rho\mu\nu} \tilde{D}_{\mu} \psi_{\nu},
\end{eqnarray}
where we have set $C=A^{-1}B^{-1/4}$, while in the bosonic sector
we have taken  $A=B^{-1/2}$ such that the covariant derivative of
the gravitino in four dimensions is given by (\ref{covder2}).

\acknowledgments{ We would like to thank J. Louis for email
correspondence, T. Mohaupt and M. Zagermann for careful reading
and suggestions of the first version of this paper.}

\end{document}